# Simultaneous High-Speed Internal and External Flow Measurements for a High-Pressure Diesel Nozzle

H. Purwar*[1], S. Idlahcen[1], L. Méès[2], C. Rozé[1], J.-B. Blaisot[1], M. Michard[2], D. Maligne[3]
[1]CORIA UMR-6614, Université de Rouen, 76801 Saint-Etienne-du-Rouvray, France
[2]LMFA UMR-5509, CNRS, École Centrale de Lyon, 69134 Écully, France
[3]DELPHI France, 41000 Blois, France
*Corresponding author: harsh.purwar@coria.fr

**Abstract**
We present an extensive experimental study focused on understanding the impact of cavitation in a high-pressure diesel nozzle on the macroscopic properties of fuel spray. Several high-speed videos of the liquid flow through a transparent, asymmetric cylindrical nozzle with a single orifice ($\phi$ = 0.35 mm) are recorded along with the videos of the resulting spray in the near-nozzle region, issued with an injection pressure of 300 bar at a frame-rate of 75 kHz. The high-repetition images of the internal flow are then used to estimate the onset of cavitation inside the transparent nozzle and the probability of development of cavitation in different regions of the nozzle with an average estimate of the amount of cavitation with time. On the other hand, recorded spray images are used to study spray penetration, cone-angles and velocity from the start of fuel injection. A novel approach is proposed for the measurement of perturbations that occur in form of big liquid structures along the spray boundary.

**Introduction**
Modern diesel injectors operate at an injection pressure as high as 2500 bar with orifice diameter of only about one hundred micrometres. As a result, there is a huge pressure variation inside these injectors, which is one of the main reasons for cavitation. In most diesel injectors the diameter of the orifice is reduced in order to increase the liquid velocity at the outlet, leading to faster fuel/air mixing. At times, due to this sudden change in the flow velocity of the fuel, the local relative pressure inside the injector nozzle drops, a result from the Bernoulli's principle. Cavitation occurs when the relative pressure drops below the vapour pressure of the fuel resulting in the formation of vapour pockets or cavities in nozzle.

A lot of efforts have been made in understanding the role of cavitation in the atomization of the fuel sprays [1-10]. C. Badock *et al.* [2] investigated cavitation phenomenon in real size diesel injection nozzles using laser light sheet and trans-illumination[1] imaging techniques. They obtained images of the internal flow and liquid jet near the nozzle exit with these two techniques and showed that using light sheet imaging the core of the flow which is covered by the cavitation films can be observed whereas with trans-illumination imaging it can't. They observed that even at high injection pressures (600 bar), the nozzle hole is filled with liquid surrounded by cavitation films. F. Payri *et al.* [5] studied the influence of cavitation on the internal flow and macroscopic behaviour of the spray (spray tip penetration and cone-angle) using two bi-orifices nozzles with different geometries -- cylindrical and convergent (or conical). They observed an increase in the spray cone-angle with the onset of cavitation. Similar results were obtained by H. Hiroyasu [3] by visualizing separately internal flow and macroscopic spray from different large-scaled nozzles. Several other researches also end up with a similar conclusion for different kinds of injector nozzles. A. Sou *et al.* [6] studied the effects of the cavitation number $\sigma$ and the Reynolds number $Re$ on cavitation in the nozzle and liquid jet using two-dimensional (or planner) nozzles (1 mm in thickness) by varying the liquid flow rates, fluid properties and nozzle sizes. They defined various cavitation regimes for cavitation in the nozzle, namely, (i) no cavitation (ii) developing cavitation (or onset of cavitation) (iii) super cavitation where the cavitation bubbles arrive close to the nozzle exit and (iv) hydraulic flip, occurs when the cavitation bubbles reach the nozzle exit and air from the surrounding is sucked in. It was shown that for 2D nozzles cavitation and liquid jet near the nozzle exit are not strongly affected by the Reynolds number $Re$ but by the cavitation number $\sigma$, which they define as follows,

$$\sigma = \frac{P_i - P_v}{0.5 \rho_L V_N^2} \quad (1)$$

here, $P_i$ is the injection pressure, $P_v$ is the vapour saturation pressure, $\rho_L$ is the liquid density, $V_N$ is the mean liquid velocity in the nozzle. A lot of other works had also been done using simple two dimensional nozzles. However, it is believed and is easy to comprehend that 2-D nozzles do not represent correctly geometric influences such as exact flow through needle seat and sac hole and, besides this, the behaviour of the cavitation may be different [2]. Recently, Z. He *et al.* [10] showed that hydraulic flip in diesel injectors depends on the length/diameter ratio using scaled-up transparent VCO nozzles. They observed hydraulic flip occurring in the nozzle with L/D ratio 4 and 6 but no hydraulic flip was observed in the nozzle with L/D ratio of 8. Hence, they concluded that hydraulic flip is less likely to occur in long nozzle diesel injectors. Similar study was conducted by H. Suh *et al.* [7] using scaled-up 2D nozzles (2 mm in thickness) for L/D ratios 1.8 and 2.7. They obtained different regimes for the two nozzles on the basis of cavitation number ($\sigma$) in which turbulence flow, cavitation and hydraulic flip were observed. Also, G. Jiang *et al.* [9] studied the cavitation density by varying various injection parameters namely, injection pressure, spray angle of the nozzle, L/D ratio, and fuel type concluding that cavitation density increases with increasing injection pressure, nozzle spray angle and saturation vapour pressure.

---
[1] At times also referred as shadowgraphy or back-lit imaging in literature.

To summarize it is now well known that cavitation up to an extent favours atomization and has been observed by several research groups as increase in the spray cone angle [5,11] and spray producing finer droplets [7]. However, it depends on a lot of factors not yet understood very well, like internal geometry of the injector, surface smoothness, fuel properties, nozzle L/D ratio, etc. On the other hand, it has also been shown that hydraulic flip might have negative effects on atomization [6]. Cavitation could also reduce the lifetime of fuel injectors due to the sudden collapse of cavitation bubbles inside them. However, this corrosion could be beneficial in cleaning the injector by removing coking deposits.

This work focuses on understanding the impact of cavitation on the macroscopic spray properties, like spray penetration, velocity and spray cone angles for the high-pressure diesel sprays in the near-nozzle region using a transparent, asymmetric cylindrical nozzle. Presentation of a method to look at the perturbations or so called "randomly occurring" big liquid structures along the spray boundary ends this article.

**Transparent nozzle design**

A transparent, asymmetric, single orifice nozzle with an orifice diameter of 350 µm was designed. The schematic of this asymmetric transparent nozzle that has been used in this work is shown in Figure 1.

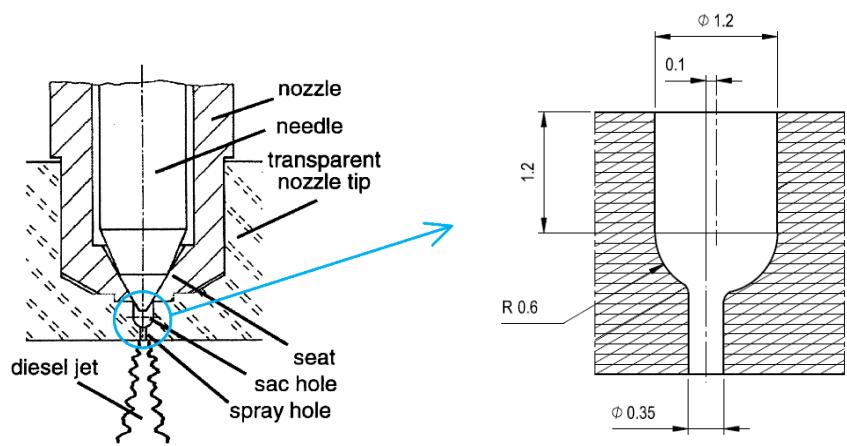

**Figure 1.** Approximate schematics for the transparent nozzle made of PMMA (left image source [2]). Distances marked on the right image are in mm.

A standard DFI1.20 injector from Delphi was trimmed at its very end (just above the orifice, under the hydraulic seat) and the single-hole transparent tip made of poly-(methyl methacrylate) (PMMA) was fixed using an intermediary metal part. Note that this intermediary metallic part may change the dynamics of the standard injector. The cylindrical sac volume ends in hemispherical shape and a single orifice which is offset with respect to the sac volume axis. This asymmetrical configuration aims to obtain an asymmetrical development of cavitation allowing a better visualization and quantification of the cavities and liquid-vapour interfaces. It ensures that the cavitation always occurs at the same side of the nozzle, irrespective of the injection conditions.

**Simultaneous high-repetition internal flow and spray measurements**

An experiment was set up to simultaneously obtain images of the internal flow (in order to visualize cavitation inside the transparent nozzle) and external flow i.e. spray (so as to study the impact of cavitation on the liquid spray atomization). A schematic of this experimental setup is shown in Figure 2. The Cavitar Cavilux Smart laser, a monochromatic (spectral width: 10 nm) incoherent pulsed illumination source centred at a wavelength of 640 nm, was being operated at a repetition rate of 75018 Hz (time between consecutive pulses: 13.32 µs) with pulse width of 10 ns. This pulse width is short enough to freeze the motion of spray at the detector for spray issued with an injection pressure of 300 bar. The incoherency of the laser source makes sure that there is no laser speckle and other artefacts in spray images arising due to the strong coherence of the traditional laser beams.

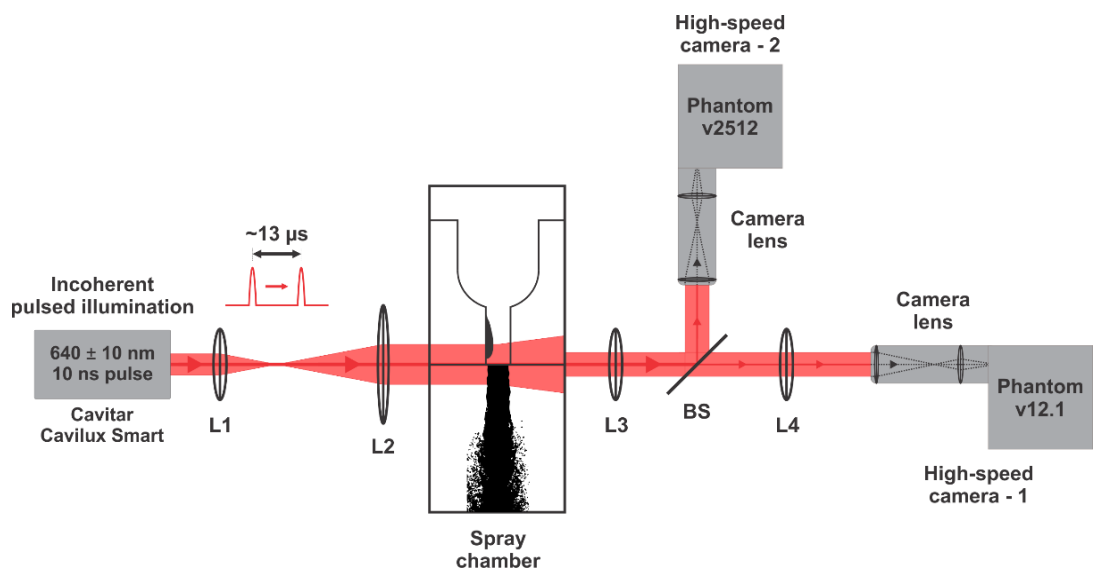

**Figure 2.** Schematic of the experimental setup for simultaneous high-speed internal flow & spray measurements.

An enlarged beam from this laser is then incident at the tip of the nozzle, illuminating both the nozzle orifice and spray in the near-nozzle region. The beam is then divided using a 50:50 beam splitter (BS) and is directed towards the two high-speed cameras for imaging. With the help of appropriate lenses thereafter, we were able to visualize simultaneously the development of cavitation inside the transparent PMMA nozzle with a magnification factor of 15.6 and spray in the near-nozzle region with a much smaller magnification of 0.97 in order to have a larger field of view (about 7.5 mm from the nozzle tip). Note that in order to visualize cavitation bubbles inside the PMMA nozzle (refractive index 1.49), an index-matched liquid obtained by mixing the diesel-like surrogate liquid[2] (Castrol calibration oil ISO 4113) with a high refractive index liquid (1-bromonapthalene) was used. Some of the relevant physical properties of this surrogate fuel are mentioned in Table 1. Note that addition of 1-bromonapthalene changes the physical properties of the calibration oil, which might lead to certain differences in spray and flow characteristics.

**Table 1.** Physical properties of diesel oil and Castrol calibration oil.

| Oil | Density (Kg/m$^3$) @15°C | Kinematic viscosity (mm$^2$/s) @40°C | Surface tension (mN/m) @20°C |
|---|---|---|---|
| Diesel oil [12,13] | ~837 | ~2.71 | ~28 |
| Castrol calibration oil | 825 | 2.53 | 27.5 |
| Calibration Oil + 1-bromonapthalene | 938 | -- | 28.6 |

The incoming laser pulses from Cavilux Smart, the two high-speed cameras shown in Figure 2 (Phantom v12.1 and Phantom v2512) and fuel injection were all synchronized using a digital delay generator from Stanford Research Systems. Several high-speed videos of the nozzle and spray in the near-nozzle region were then recorded at a frame rate of 75 kHz with reduced region of interest on both cameras. The spray was issued inside a closed chamber, to avoid any damage to the optical devices in case of a leak. The chamber was maintained at the atmospheric temperature and pressure (ATP). The injection duration was 2.5 ms.

An example of the images obtained using the described optical setup (Figure 2) for internal and external flow (spray) are shown in Figure 3 for two different time-stamps after injector activation for the same injection event. All time references in this article are with respect to the start of activation of injector (SOAI), which is about 0.3 ms before the actual start of fuel injection (SOI).

As a consequence of the refractive index matching, refraction of light at the nozzle-liquid boundaries is not visible and the liquid phase is bright in the nozzle images, whereas vapour or cavitation pockets are dark due to the interaction of light with the liquid-vapour interfaces. On the other hand, for the spray images liquid is dark and the background (air) is bright.

---

[2] Used in order to avoid any accidental combustion and reduce risks for experimentalists.

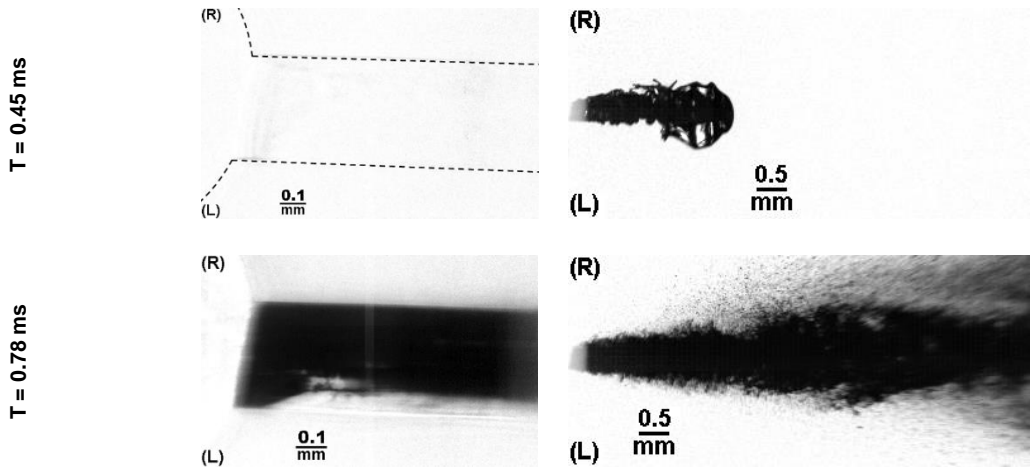

**Figure 3.** An example of obtained nozzle and spray images after T milliseconds from the SOAI for the same injection event. (L) and (R) indicates left and right sides respectively in spray and nozzle images.

**Injector needle-lift and laser timings**

In order to study the impact of cavitation on the atomization of spray, it is important to know exactly when, after the start of activation of injector, does the cavitation start to develop in the injector nozzle. For this reason, the DFI1.20 Delphi injector was fitted with a non-contact, eddy current based displacement measurement system from Micro-epsilon with a sensor suitable for targets made of ferromagnetic electrically conductive materials, such as this injector, to measure injector needle lifts after the SOAI. High-frequency alternating current flows through a coil cast in a sensor casing. The electromagnetic coil field induces eddy currents in the conductive target (injector needle) thus changing the AC resistance of the coil. This change in the impedance is interrupted by demodulation electronics, which generates an electrical signal proportional to the distance of the target from the sensor. This electrical signal can then be easily measured using an oscilloscope or any other similar equipment used for such data acquisitions. At the fluid pressures used in this work, no additional correction of the measure was needed to improve the precision of the lift measurement. The specific current pulse requirement for the injector (supplied by the Delphi team) that is used in this work was fulfilled by deploying the IPoD coil (EFS 8427). By changing specific parameters of this device, desired current pulse was obtained. This current pulse was measured using a Hall-effect based current clamp E3N from Chauvin Arnoux. The output signal from the current clamp is directly proportional in form and amplitude to the DC and AC components of the current measured. Figure 4 shows this very specific current pulse along with the injector needle lift, both averaged over 50 fuel injections. The timings of laser pulses used for recording cavitation and spray images are marked in red in this figure. All these parameters were measured simultaneously using high-speed data acquisition board, DT9836 from Data Translation.

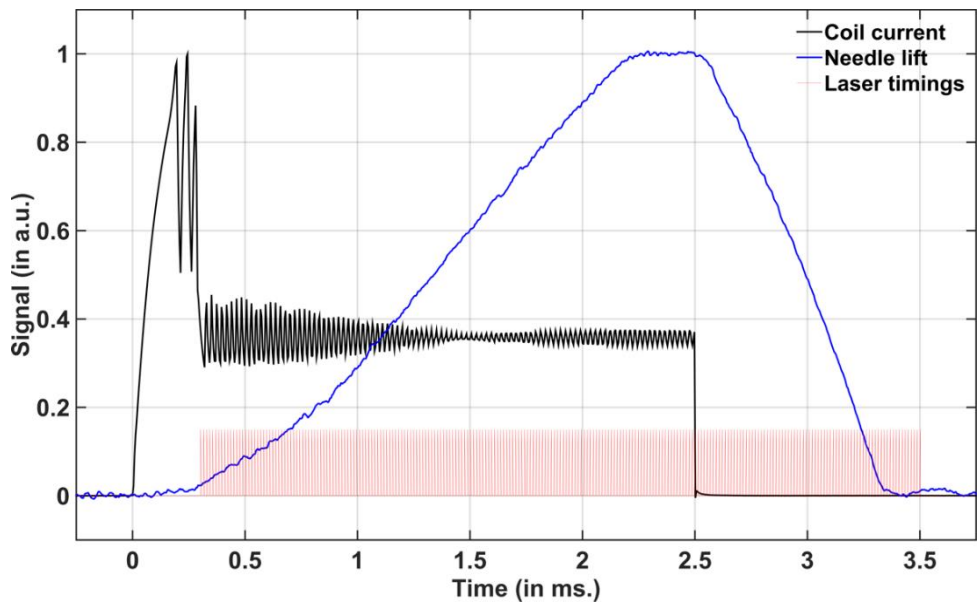

**Figure 4.** Injector coil current pulse, needle lift measurement and laser timings. The injection duration was 2.5 ms. SOI about 0.3 ms after the SOAI (i.e. at T = 0).

**Results and discussion**
After simultaneous acquisition of high-speed videos of cavitation inside the transparent nozzle and resulting spray in the near-nozzle region for 100 fuel injection events, the following characteristics were estimated from these videos/images. Also, in order to quantify cavitation, a quantity named cavitation probability is estimated from the images of the internal flow and is explained in the following subsections.

Shape of the jet
The spray images at the onset of fuel injection (mushroom-like liquid structure, Figure 3) look very similar to the high-resolution diesel spray images obtained at 400 bar by C. Crua *et al.* [14], where they used long-range microscopy with SIM-16 high-speed camera from Specialised Imaging to investigate the primary atomization of diesel fuel in the near-nozzle region. They concluded that the fluid that first exited the nozzle resembled mushroom-like structures due to the residual fluid trapped in the sac and orifice between injections. Such structures were not only observed for atmospheric but also for realistic engine conditions by other researchers as well [15].

Spray penetration length
The recorded spray images were used to estimate the penetration length of the jet as shown in Figure 5. The tip (end point) of the jet was traced in the consecutive spray images to calculate its displacement with time.

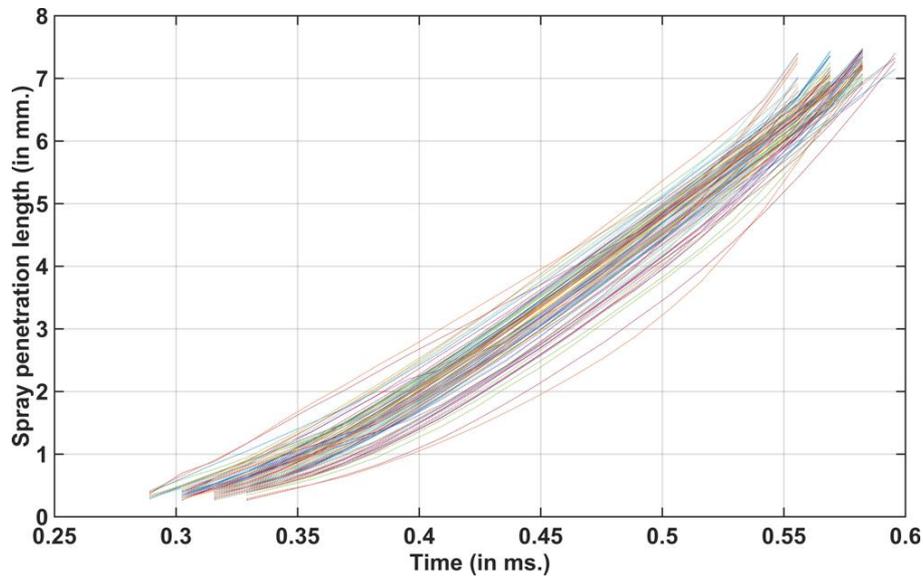

**Figure 5.** Evolution of penetration length with time for the fuel spray ejected in a chamber maintained at ATP with an injection pressure of 300 bar.

In Figure 5 the penetration length (or displacement) in the near-nozzle region is shown for 100 different fuel injections with same injection parameters. Recently, C. Crua *et al.* [14] published their findings with the spray tip penetration in the near-field of the nozzle at an injection pressure of 400 bar. Their measurements, though limited by the number of frames captured by the SIM-16 high-speed camera that was used for recording the high speed video of the fuel spray, show that the overall behaviour of the spray tip penetration at 400 bar of injection pressure is very similar to our results shown in Figure 5 at an injection pressure of 300 bar. Note that due to about 0.04 ms jitter in the fuel injection system, there is about 0.04 ms spread in the estimated penetration lengths for different fuel injections.

Cavitation probability and injector needle-lift
The average cavitation video/images from a sample size of 100 fuel injections was generated by simply taking an average of the recorded cavitation images after background subtraction for different fixed laser timings shown in Figure 4. The average cavitation images represent the probability of occurrence of vapour pockets inside the transparent nozzle. Hence, in the average cavitation images, if a region is dark (black), higher is the probability of cavitation formation in this region. These average cavitation images were also used for the calculation of cavitation probability in order to quantify the amount of cavitation at a particular instant of time after the SOAI. We define cavitation probability ($CP$) as,

$$CP = \frac{1}{N} \sum_{i,j} \left(1 - \frac{\bar{C}_{i,j}}{255}\right) \qquad (2)$$

Here, $\bar{C}_{i,j}$ represents value of the $(i,j)^{th}$ pixel in the 8-bit greyscale average cavitation image ($\bar{C}_{i,j}/255 = 0 \Rightarrow$ cavitation, $\bar{C}_{i,j}/255 = 1 \Rightarrow$ no cavitation). Thus, $(1 - \bar{C}_{i,j}/255)$ operation represents image inversion, 255 being the maximum value that each image pixel can have in an 8-bit greyscale image. The summation is carried out over all the pixels in the region that includes the cavitating nozzle with $N$ being the total number of pixels in this region. Figure 6 shows an example of the average cavitation image for T = 0.78 ms after the SOAI. The pixels that are summed up to obtain $CP$ are inside the marked polygon (with red solid lines) in this figure.

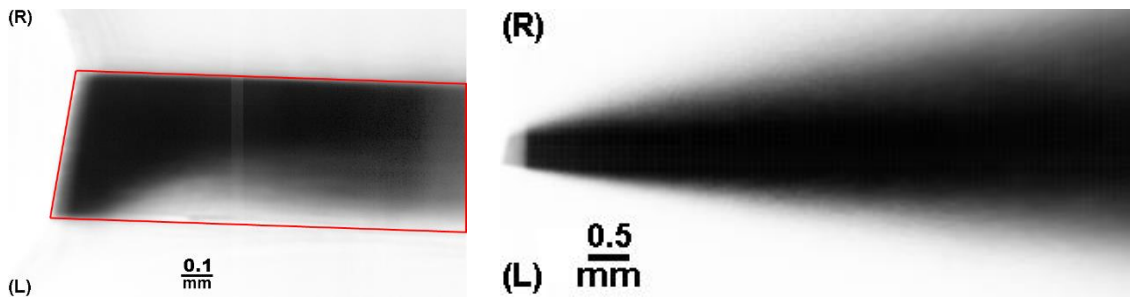

**Figure 6.** An example of the average cavitation (left) and spray (right) image (averaged over 100 injection events) for T = 0.78 ms after the SOAI. The pixels (or area) surrounded by the red lines are summed up for the calculation of $CP$.

The variation of cavitation probability ($CP$) estimated from the average cavitation images is shown in Figure 7 with time after the SOAI.

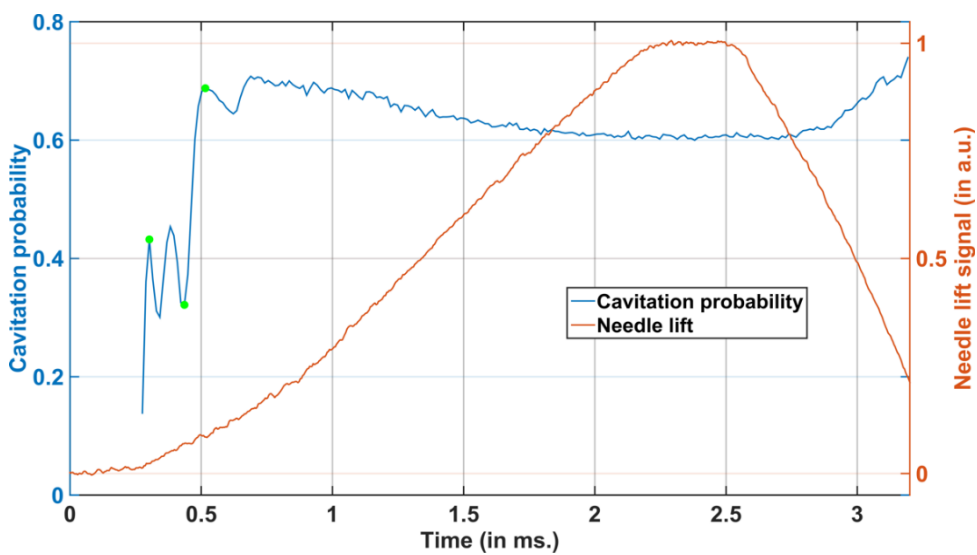

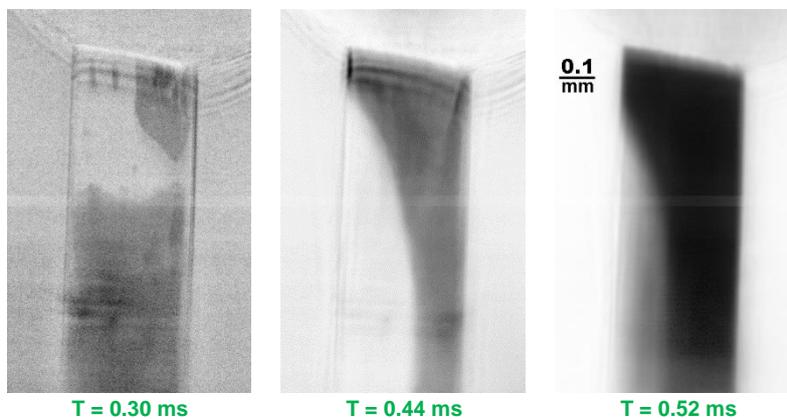

T = 0.30 ms    T = 0.44 ms    T = 0.52 ms

**Figure 7.** Top: Cavitation probability ($CP$) for different times after the SOAI. Needle lift is also plotted along with $CP$ (same as shown in Figure 4) for a better understanding of the onset of cavitation with needle lift. Bottom: Average cavitation images for three time stamps marked with green dots in cavitation probability plot.

It is very clear from Figure 7 that cavitation inside the asymmetric transparent nozzle starts to develop at a very early stage of the fuel injection. The amount of cavitation increases rapidly for the first 0.5 ms after the SOAI and then slowly stabilizes to 60% of the maximum possible value until the complete needle lift. For a better understanding of the onset of cavitation in nozzle, average cavitation images for three different time stamps are also shown in Figure 7. At T = 0.30 ms we see a small peak in the cavitation probability curve, however if we look at the corresponding average cavitation image for this time, we can see that the cavitation is not very stable at this time as the grey level of cavitation pockets is not very dark.

Spray cone angle

There could be several ways of estimating spray cone angles from the spray images recorded by the high-speed camera 1 in Figure 2, depending upon how these images are binarized [16]. We use the following approach to estimate the spray cone angles. First, the entire dataset of the normalized spray images was used to form an average normalized spray image (one image for each time step). This average normalized spray image was then used to find an appropriate threshold for binarization using the algorithm proposed by N. Otsu [17]. Individual spray images were then binarized using this threshold value (same value for all the normalized images in the dataset for a particular time-step). After binarization, the core jet was separated from the droplet clouds and any small disconnected spray structure or ligament. Next, each row of this binary core jet image was scanned in the direction of flow to locate the extremities of the core jet, giving us the two spray boundaries. Both the left and the right side spray boundaries were then fitted with straight lines and the angles made by these fitted lines with the nozzle hole axis were computed. An average of these estimated half-cone angles was then made over the entire spray dataset (100 fuel injections) for each time-step. The evolution of spray half-cone angles with time after SOAI is shown in Figure 8 separately for the two sides of the spray.

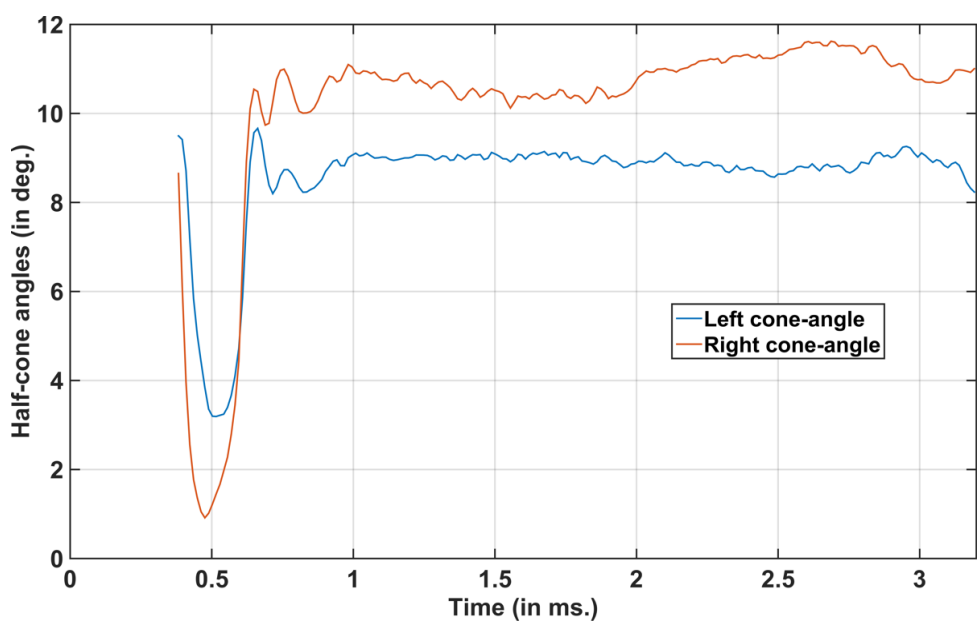

**Figure 8.** Average left and right spray cone angles computed using the method described above for each time step after SOI (corresponds to different injector needle lifts and cavitation probability).

It can be observed that there is a significant difference between the left and the right side angles for the recorded spray images, which clearly is a consequence of the asymmetric nozzle design. Cavitation is more developed on the right side of the nozzle clearly visible in the cavitation images (see Figure 3 or Figure 7). The overall higher value of the right side half-cone angle indicates that the spray is more developed on the right side (or cavitating side) than on the left (non-cavitating side). It can as well be seen on the average spray image shown in Figure 6 for a particular time step. Note that the sudden decrease and increase in the spray cone angle at the start of fuel injection is due to the propagation of the mushroom-shaped liquid structure. The cone angles are more or less stable thereafter.

Spray velocity

To fully describe the motion of any object in general, one must define its position and velocity with time. A spray image at any particular instant of time after the SOI gives its position in time. An average over a significant number of injections would give the probability of occurrence of liquid or droplets in space. The behaviour of the fuel spray is very complex and non-redundant at high injection pressures and we want an approach which could take this uncertainty into account and produce a statistical mapping of the velocity vectors. Such an approach for velocity computation briefly described below, is based on calculation of correlation between two consecutive images recorded within a very short period of time proposed by D. Sedarsky *et al.* [18].

Two consecutive back-lit spray images are used here to calculate spray velocity. The prerequisite for this method to work is that the two spray images should be recorded within a very short period of time so that the spray structure remains almost intact in the two images and yet the spray moves slightly. The spray images recorded using the high-speed camera at 300 bar of injection pressure with an inter-frame delay time of 13.32 µs were normalized and paired (1st & 2nd, 2nd & 3rd and so on). These were then subjected to the cross-correlation based velocity computation. A small section on the first image near the edge of the spray was chosen and then the similar structures and ligaments were searched in the second image of the pair. The displacement of the entire chosen section was then estimated from the maximum value of the normalized cross-correlation. Note that usually a very small section on the first image of the image-pair is chosen and hence the estimated value of displacement is justifiable. Since for these injection parameters the spray does not change too much, at least not in the beginning of the injection, the displacement vectors were found with high cross-correlation values. Only vectors with normalized cross-correlation values greater than 0.80 were selected and the rest were rejected. Choosing a low threshold value for normalized cross-correlation is more likely to give erroneous displacement vectors and if a very high value is chosen one might not find sufficient vectors. We chose this particular value by hit and trial and it was made sure that there are no erroneous displacement vectors in the results. For demonstration, the selected displacement vectors for two such consecutive images are shown in Figure 9.

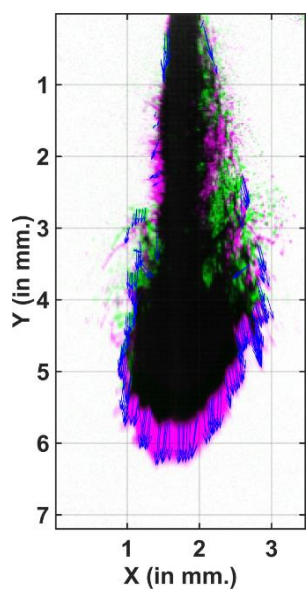

**Figure 9.** Displacement vectors computed by correlating the two spray images obtained within a short period of time (13.32 µs). The first image of the pair is shown in green, the latter in magenta and the overlapping regions are shown in black. Displacement vectors are shown in blue.

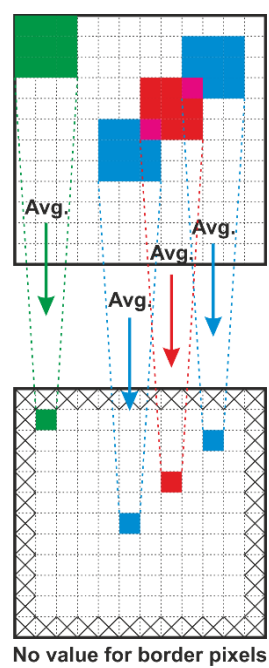

**Figure 10.** Schematic to emphasize on the moving average procedure used to generate velocity maps.

The time for this displacement is same as the inter-frame time for the camera. Using this information, the velocities of the liquid structures and ligaments visible on the spray images were estimated.

The velocity vectors for all the consecutive image-pairs at the beginning of the fuel injection (i.e. until 0.66 ms. after SOAI) were computed. All the velocity vectors within a short duration of time were grouped together according to their spatial locations. This was done for 100 such injections and an average velocity map was generated. A moving average over all the vectors inside a 3 x 3 pixel box was then taken in order to smoothen out the velocity maps (see the schematic in Figure 10).

The evolution of the average spray velocity with time just after SOI is shown in Figure 11.

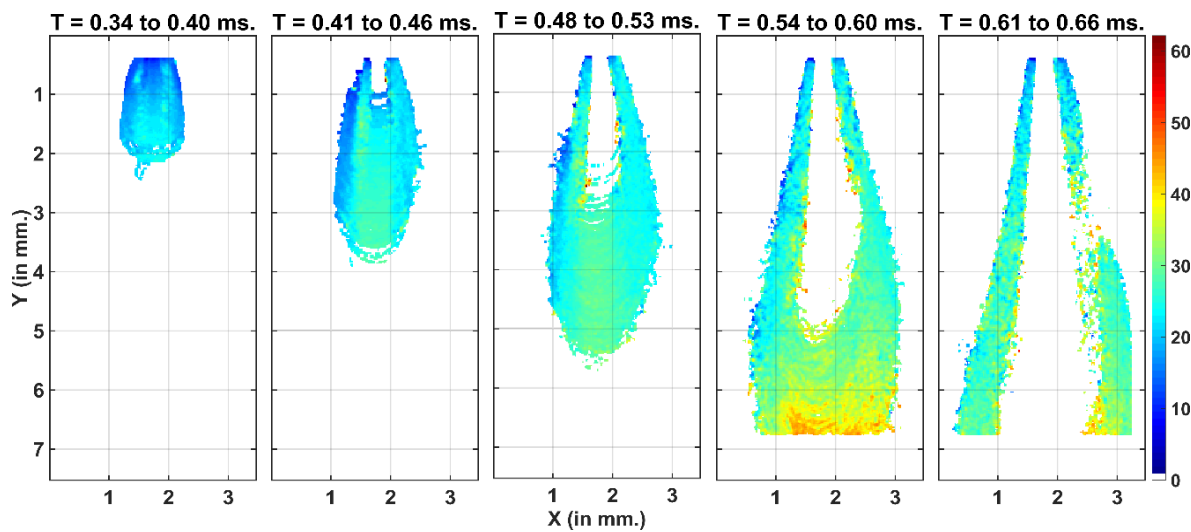

**Figure 11.** Average spray velocity maps computed using consecutive spray images captured with the high-speed camera at 300 bar of injection pressure. The colour bar indicates velocity in m/s. T is the time with respect to the SOAI.

As we mentioned before, in order for this method to work, we need spray structure to be almost intact from one frame to another. At the start of fuel injection, the spray velocity is not very high and this condition is satisfied for an inter-frame delay time of 13 µs. However, as fuel injection continues the spray velocity increases and as we can see in the right most velocity map (for T = 0.61 to 0.66 ms.) in Figure 11, we don't have a lot of velocity points along the spray edge. For this reason, we must reduce the inter-fame delay by using a higher repetition camera, such as a PIV camera. Not to forget that multiple scattering also plays an important role here. Due to the formation of intense droplet cloud surrounding the spray after a few ms from the SOI, visibility of spray structures is reduced leading to only a handful of displacement vectors or even erroneous vectors.

H. Purwar *et. al* [19] showed that the spray velocity on the non-cavitating side of the nozzle is significantly higher than the spray velocity on the cavitating side using this method on spray images obtained using a PIV camera. This behaviour in the spray velocity however, is not observed here at the very start of fuel injection. This is because cavitation is not yet properly developed inside the nozzle and whatever cavitation is there at a later time, for example after 0.5 ms (see Figure 7), high velocity of the jet does not allow the calculation of a lot of displacement vectors with high correlation value. For this reason, we present another approach to look at the spray velocities along the spray edges.

Off-boundary displacements
We know from the calculation of the average spray cone angles (Figure 8) that their values are stable after 1.0 ms from the activation of injector. This means that we could calculate the time-average spray boundary and look at the perturbations along this boundary in time. This was done by first generating two time-average spray images, one for the left side and other for the right, starting after formation and stabilization of the fuel spray i.e. after T = 1.0 ms from the SOAI by averaging over all the greyscale images (151 in number) recorded within this time duration for a particular fuel injection event. The time-average images were then binarized using a threshold value determined using Otsu's method and the edges of the spray were determined as shown with red solid line curves in Figure 12 along with the time-average spray images for the left and right sides of the spray. The threshold values for binarization for both sides of the sprays were chosen such that the extracted spray edges after binarization could be a close approximation to the boundary of the time-average spray images. To summarize, the red solid curve shown in Figure 12 will now be treated as the boundary of the spray at all times for this particular fuel injection event. Note that the average boundary of the spray is not a straight line here but a curve instead.

Next, the individual spray images at different times starting from T = 1.0 ms to T = 3.20 ms were binarized using the same threshold value that was used to binarize the time-average spray images and the displacement of the off-boundary structures along the horizontal or radial direction with respect to the time-average spray boundary was determined. To do so, the predetermined spray boundaries from the time-average spray images for both sides were considered as the boundaries of the spray for individual images at each time-step. Figure 13 shows a few of such individual spray images at different times with the time-average spray boundary (in red) and points out a big off-boundary liquid structure which moves and spreads as the spray progresses in time.

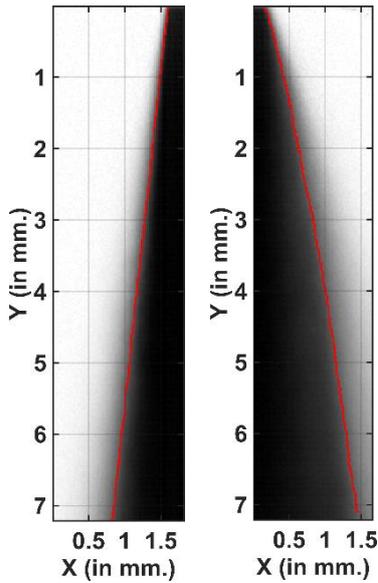 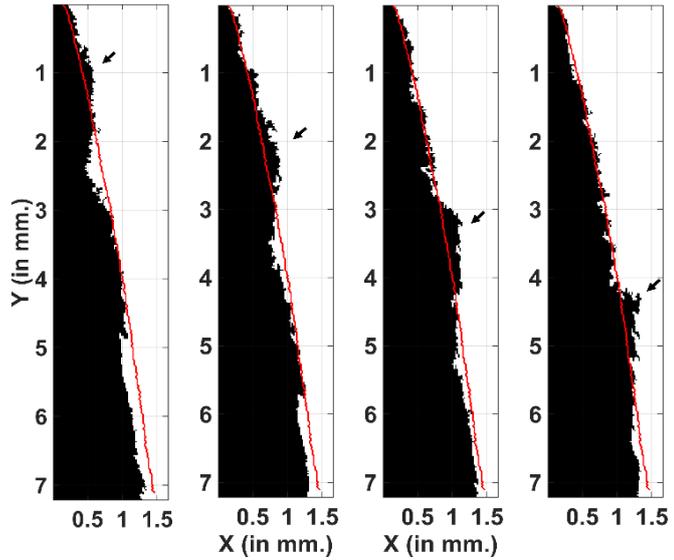

**Figure 12.** Time-average fuel spray images for left and right sides of the spray with average spray boundary (in red) for a particular injection event.

**Figure 13.** Occurrence of big liquid structures along the boundary of the spray, their movement and spreading in time. Time is increasing from left to right in steps of 13.32 µs.

The displacement of such off-boundary structures in time is determined in this subsection. If liquid structures move away from the spray, like the structure pointed out in Figure 13, their displacement was considered as positive and in case it is the time-average spray boundary that exceeds the limits of the spray in individual spray images, the displacement was considered as negative. Note that ideally one should determine the off-boundary displacement by measuring the perpendicular distance of the off-boundary structure from the spray boundary. However, for simplicity and since the cone angle is small, here the off-boundary displacement is just the horizontal distance of the off-boundary structure from the spray boundary for each boundary pixel. The schematic in Figure 14 addresses this point. Moreover, at times it might be difficult to determine or even define the displacement of the liquid structures due to their complicated shapes. For example, the middle structure shown in Figure 14 could have two values of displacements while the bottom structure could have both negative and positive displacements. Here for both these cases the maximum value of the displacement was considered. The negative displacement was always ignored even if the magnitude of negative displacement was more than that of the positive displacement.

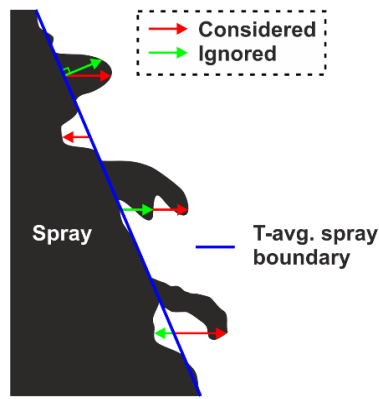

**Figure 14.** Ambiguous off-boundary structures.

The off-boundary displacement plots for left and right side boundaries are shown separately in Figure 15.

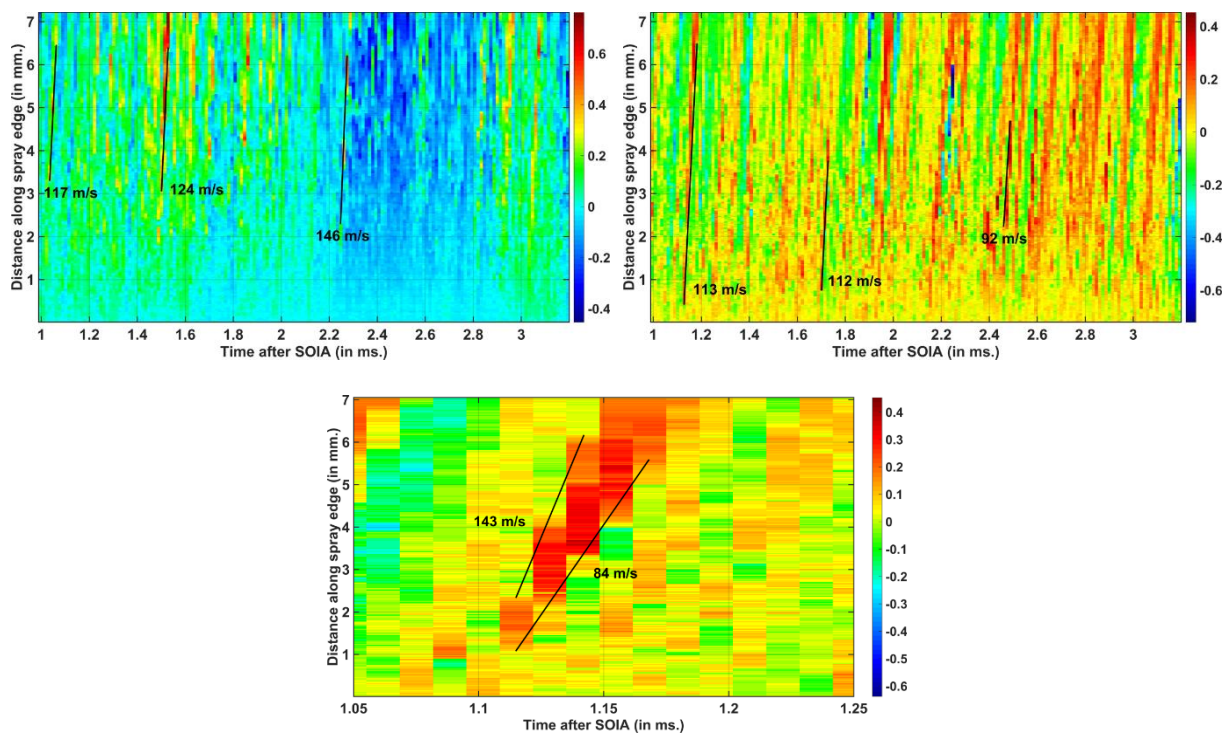

**Figure 15.** Top: Off-boundary displacement plots for left (left) and right (right) sides of the spray for a particular injection event. Bottom: Magnified view to emphasize on the spreading of liquid structures. The colour bar shows displacement in millimetres in horizontal (or radial) direction.

The slopes of the slanted lines represent the speed of big structures in the off-boundary displacement plots at different times after the SOI as demonstrated in Figure 15. From these slopes, it can be noted that in general, the velocity of the liquid, or at least of the big liquid structures, is larger on the left side (non-cavitating) of the spray than that on the right (cavitating). This liquid deformation representation opens a new way for statistical analysis of the jet edge complexity. Another characteristic which can be extracted from these off-boundary displacement plots is the spreading of the edge perturbations. In some cases, we can see that these slanted lines spread as we move further away from the nozzle in the direction of the spray. This would mean that there is not just a particular velocity for each of these structures but instead, a velocity gradient which would allow the spreading of the structures as the spray progresses in time (see the magnified view in Figure 15).

These boundary perturbations could be a result of any or all of the ongoing processes inside the fuel injector during injection, for example, needle vibrations, turbulence, cavitation, etc. Here, it should be noted that these plots were just for one particular injection event. However, this approach could easily be transformed to give statistically significant results in future and perhaps these minor perturbations along the spray boundary could give us an insight about what is going on inside the injector.

**Summary and Conclusions**

The aim of this work was to study the impact of cavitation inside the high-pressure diesel nozzle on the atomization of sprays. A transparent, cylindrical single-hole nozzle ($\phi$ = 0.35 mm) was designed and manufactured using PMMA with non-overlapping sac and nozzle axes. This asymmetric design of the nozzle leads to an asymmetric development of cavitation inside the nozzle after about 0.45 ms from the start of activation of injector.

An experiment aimed at recording simultaneously, images of internal and external flow through the nozzle at a high-repetition rate was set up. Several cavitation and spray images were recorded simultaneously with different magnifications at 75 kHz. The images of the internal flow through nozzle were used to study the onset of cavitation and the probability of cavitation in different regions of the transparent nozzle by averaging over 100 fuel injections. Similarly, spray images were used to calculate spray penetration in the near nozzle region (within 7.5 mm from the nozzle tip), spray cone angle over the entire injection duration and average spray velocity at the SOI. From these characterizations it can be concluded that the spray cone angle is positively correlated with the cavitation inside the nozzle. In the near nozzle region and at the very start of the fuel injection (within 0.66 ms from SOAI) the average spray velocity on the two sides (cavitating and non-cavitating) is the same. We showed in our previous experiments that the spray velocity after the onset of cavitation is very different on these two sides. The velocity at the cavitating side is much smaller than that on the non-cavitating side, which suggests that the atomization time-scales on the two sides of the nozzle are different [19].

Towards the end, we demonstrate a novel approach to look at the big-liquid pockets that appear along the spray boundary from time to time during fuel injection. For a particular injection event, we plotted off-boundary displacements for the cavitating and non-cavitating sides and a very clear asymmetry has been observed in these displacement plots. If on one side fuel spray is atomizing faster, one would expect to observe more perturbations along the spray boundary, in the form of big-liquid pockets moving away from the spray core. This will show up as a lot of fluctuations in the displacement plots. And indeed, we observe the same in our case with asymmetric development of cavitation inside the nozzle, which ones again suggest that the atomization time-scales are different on the two sides of the spray. Moreover, these plots can be used to calculate the velocity of these perturbations along the boundary at various time steps after the SOI by simply looking at the slopes of the slanted linear structures in the displacement plots. Such an approach could be very helpful in understanding various coupled phenomenon simultaneously occurring inside the high-pressure injectors and could open up a new way of characterizing liquid fuel sprays.

**Acknowledgements**
Authors acknowledge financial support from the French National Research Agency (ANR) through the project CANNEx (ANR-13-TDMO-0003).

**References**
[1] He, L., and Ruiz, F., 1995, Atomization and Sprays, 5(6), pp. 569-584.
[2] Badock, C., Wirth, R., Fath, A., and Leipertz, A., 1999, International Journal of Heat and Fluid Flow, 20(5), pp. 538-544.
[3] Hiroyasu, H., 2000, Atomization and Sprays, 10(3-5), pp. 511–527.
[4] Tamaki, N., Shimizu, M., and Hiroyasu, H., 2001, Atomization and Sprays, 11(2), pp. 14.
[5] Payri, F., Bermúdez, V., Payri, R., and Salvador, F. J., 2004, Fuel, 83(4-5), pp. 419-431.
[6] Sou, A., Hosokawa, S., and Tomiyama, A., 2007, International Journal of Heat and Mass Transfer, 50(17-18), pp. 3575-3582.
[7] Suh, H. K., and Lee, C. S., 2008, International Journal of Heat and Fluid Flow, 29(4), pp. 1001-1009.
[8] Desantes, J. M., Payri, R., Salvador, F. J., and De La Morena, J., 2010, Fuel, 89(10), pp. 3033-3041.
[9] Jiang, G., Zhang, Y., Wen, H., and Xiao, G., 2015, Energy Conversion and Management, 103, pp. 208-217.
[10] He, Z., Shao, Z., Wang, Q., Zhong, W., and Tao, X., 2015, Experimental Thermal and Fluid Science, 60, pp. 252-262.
[11] Zhang, L., Gao, Y., Li, L. G., Deng, J., Gong, H. F., and Wu, Z. J., Aug. 23-27 2015, 13th International Conference on Liquid Atomization and Spray Systems.
[12] Alptekin, E., and Canakci, M., 2009, Fuel, 88(1), pp. 75-80.
[13] Schaschke, C., Fletcher, I., and Glen, N., 2013, Processes, 1(2), pp. 30-48.
[14] Crua, C., Heikal, M. R., and Gold, M. R., 2015, Fuel, 157, pp. 140-150.
[15] Badock, C., Wirth, R., Fath, A., and Leipertz, A., 1999, International Journal of Heat and Fluid Flow, 20(5), pp. 538-544.
[16] Purwar, H., 2015, "Ultrafast imaging of fuel sprays: development of optical diagnostics, image processing". Université de Rouen, France.
[17] Otsu, N., 1979, IEEE Transactions on Systems, Man, and Cybernetics, 9(1), pp. 62-66.
[18] Sedarsky, D., Idlahcen, S., Rozé, C., and Blaisot, J.-B., 2013, Experiments in Fluids, 54(2), pp. 1451.
[19] Purwar, H., Lounnaci, K., Idlahcen, S., Rozé, C., Blaisot, J., Méès, L., and Michard, M., Aug. 23-27 2015, 13th International Conference on Liquid Atomization and Spray Systems.